\documentstyle[manuscript,aps]{revtex}
\begin{document}

\title{Two-Point Green's Function in ${\cal PT}$-Symmetric Theories}

\author{Carl M. Bender$^1$, Stefan Boettcher$^2$, Peter N. Meisinger$^1$, and Qinghai Wang$^1$}

\address{${}^1$Department of Physics, Washington University, St. Louis, Missouri 63130, USA}

\address{${}^2$Department of Physics, Emory University, Atlanta, Georgia 30322, USA}

\date{\today}

\maketitle

\begin{abstract}
The Hamiltonian $H={1\over2} p^2+{1\over2}m^2x^2+gx^2(ix)^\delta$ with $\delta,g\geq0$ is non-Hermitian, but the energy levels are real and positive as a consequence of ${\cal PT}$ symmetry. The quantum mechanical theory described by $H$ is treated as a one-dimensional Euclidean quantum field theory. The two-point Green's function for this theory is investigated using perturbative and numerical techniques. The K\"allen-Lehmann representation for the Green's function is constructed, and it is shown that by virtue of ${\cal PT}$ symmetry the Green's function is entirely real. While the wave-function renormalization constant $Z$ cannot be interpreted as a conventional probability, it still obeys a normalization determined by the commutation relations of the field. This provides strong evidence that the eigenfunctions of the Hamiltonian are complete. \end{abstract}

\vspace{.3in}
In this Letter we study the quantum mechanical theory described by the Hamiltonian 
\begin{equation} H={1\over2}p^2+{1\over2}m^2x^2+gx^2(ix)^\delta\qquad(\delta,g\geq0).
\label{e2}
\end{equation}
We treat this theory as a one-dimensional self-interacting scalar Euclidean quantum field theory whose Hamiltonian is \begin{eqnarray} H={1\over2}(\nabla\phi)^2+{1\over2}m^2\phi^2+g\phi^2(i\phi)^\delta\qquad(\delta,
g\geq0)
\label{e1}
\end{eqnarray}
and study the two-point Green's function for this theory.

The quantum mechanical theory in (\ref{e2})
has been studied in great detail in the past few
years because, while it is not Hermitian, it does possess ${\cal PT}$ symmetry. (Under parity $\cal P$, $x\to-x$ and $p\to-p$; under time reversal $\cal T$, $x \to x$, $p\to-p$, and $i\to-i$. Thus, the operator ${\cal PT}$ commutes with the Hamiltonian $H$.) It appears that, as a consequence, all of the energy levels are real and positive\cite{X1,X2,X3,X4,X5,X6,X7,X8,X9,X10,X11,X12,X13,X14}. The first rigorous proof of the reality of the spectrum for the massless case $m=0$ is given in Ref.~\cite{X13}. Direct numerical evidence for the reality and positivity of the spectrum can be found by performing a Runge-Kutta integration of the associated Schr\"odinger equation \cite{X1}. Alternatively, the large-energy eigenvalues of the spectrum can be calculated with great accuracy by using WKB techniques \cite{X1,X3,BO}. One can also demonstrate the reality and positivity of the spectrum by calculating exactly the spectral zeta function (the sum of the reciprocals of the eigenvalues). This was done by Mezincescu \cite{X7} and Bender and Wang \cite{X9}.

Non-Hermitian quantum mechanical ${\cal PT}$-symmetric theories have been studied extensively, but not much work has been done on the corresponding scalar quantum field theories. Three major studies have been published. The first \cite{Sa} shows how to solve the Schwinger-Dyson equations for such theories. The interesting result is that the beta functions for the $ig\phi^3$ and the $-g\phi^4$ theories have the opposite signs from those of the conventional $g\phi^3$ and $g\phi^4$ theories. Thus, in four dimensional space these non-Hermitian theories are asymptotically free and are nontrivial. The second paper \cite{Ya} is a nonperturbative calculation in $D$-dimensional space of the one-point Green's function $G_1$ in such theories. It is quite remarkable that $G_1=\langle0|\phi(0)|0\rangle$ is nonvanishing for all $\delta>0$. This is true even for the case $\delta=2$, which corresponds to a $-g\phi^4$ field theory. Hence, this model might possibly be useful to describe the physics of the Higgs sector. The third paper \cite{Jo} concerns bound states in such theories. It is shown that for sufficiently small coupling constant, a $-g\phi^4$ theory possesses $n$-particle bound states for any $n$. In fact, there is a range of $g$ for which there are just two-particle and three-particle bound states but not bound states of four or more particles, which is somewhat reminiscent of the way in which quarks form bound states.

In this Letter we consider the field theoretic connected two-point Green's function, which in Minkowski space is defined in general by \begin{equation} G_2^{\rm Minkowski}(x-y)=\langle0|T[\phi(x)\phi(y)]|0\rangle_{\rm connected}, \label{e3} \end{equation} where $T$ represents time ordering. To discuss this equation in the context of quantum mechanics, which is equivalent to quantum field theory in one time and no space dimensions, we incorporate the time ordering by using the Heaviside step function $\theta(x)$: \begin{equation} \langle0|T[\phi(x)\phi(y)]|0\rangle=\langle0|\phi(x)\phi(y)|0\rangle\theta(x-y)+
\langle0|\phi(y)\phi(x)|0\rangle\theta(y-x).
\label{e4}
\end{equation}
We evaluate $\langle0|\phi(x)\phi(y)|0\rangle$ by inserting a complete set of
states:
\begin{eqnarray}
\langle0|\phi(x)\phi(y)|0\rangle &=&
\sum_{n=0}^\infty \langle0|\phi(x)|n\rangle\langle n|\phi(y)|0\rangle\nonumber\\ &=& \sum_{n=0}^\infty \langle0|e^{-iHx}\phi(0)e^{iHx}|n\rangle
\langle n|e^{-iHy}\phi(0)e^{iHy}|0\rangle \nonumber\\
&=& \sum_{n=0}^\infty [\langle0|\phi(0)|n\rangle]^2 e^{i(E_n-E_0)(x-y)}. \label{e5} \end{eqnarray} Thus, if we include the effect of time ordering, we obtain \begin{equation} \langle0|T[\phi(x)\phi(y)]|0\rangle=
\sum_{n=0}^\infty [\langle0|\phi(0)|n\rangle]^2 e^{i(E_n-E_0)|x-y|}. \label{e6} \end{equation} In this sum the $n=0$ term $[\langle0|\phi(0)|0\rangle]^2$, which is the square of the one-point Green's function, represents the disconnected part of the two-point Green's function. Thus, the connected two-point Minkowski space Green's function is given by \begin{equation} G_2^{\rm Minkowski}(x-y) = \sum_{n=1}^\infty [\langle0|\phi(0)|n\rangle]^2 e^{i(E_n-E_0)|x-y|}. \label{e7} \end{equation}

In Euclidean space we replace the $i$ in the exponent of (\ref{e7}) by $-1$ and get \begin{equation} G_2(x-y)=\sum_{n=1}^\infty [\langle0|\phi(0)|n\rangle]^2 e^{-(E_n-E_0)|x-y|}. \label{e8} \end{equation} We perform a Fourier transform to find the Green's function in momentum space: \begin{equation} G_2(E)=\sum_{n=1}^\infty{2(E_n-E_0)[\langle0|\phi(0)|n\rangle]^2\over
E^2+(E_n-E_0)^2},
\label{e9}
\end{equation}
where we use the identity ${2\mu\over E^2+\mu^2}=\int_{-\infty}^\infty dx\, e^{-|x|\mu}e^{iEx}$.

The difference of the first-excited-state energy $E_1$ and the ground-state energy $E_0$ is defined as the renormalized mass $M$. It is traditional to define the wave-function renormalization constant $Z$ as the coefficient of the one-particle pole ${1\over E^2+M^2}$. In quantum mechanics we have an infinite sum of poles instead of a cut. Thus, we define the $n$th wave-function renormalization constant $Z_n$ by \begin{equation} Z_n\equiv2(E_n-E_0)[\langle0|\phi(0)|n\rangle]^2.
\label{e11}
\end{equation}
Finally, we express the matrix elements above in terms of
the usual coordinate-space energy eigenfunctions $\psi_n(x)$, which satisfy the time-independent Schr\"odinger equation \begin{equation} H\psi_n(x)=E_n \psi_n(x). \label{e12} \end{equation} We thus obtain a representation for $Z_n$ as a ratio of integrals: \begin{equation} Z_n = {2(E_n-E_0)\left[\int dx\,x\psi_0(x)\psi_n(x)\right]^2\over
\left[\int dx\,\psi_0^2(x)\right]\left[\int dx\,\psi_n^2(x)\right]}. \label{e13} \end{equation} We have not seen this general result before in the literature.

We now make a crucial observation. For ordinary real Hamiltonians the solution to the Schr\"odinger equation (\ref{e12}) is real, so it is not necessary in
(\ref{e13}) to use complex conjugates of wave functions. That is, we have written $\psi_n^2(x)$ and not $|\psi_n(x)|^2$. In our upcoming discussion of the complex Hamiltonian (\ref{e2}) we will continue to use the formula (\ref{e13}). It would be wrong to use complex conjugates of wave functions because for non-Hermitian ${\cal PT}$-symmetric Hamiltonians the statement of orthogonality 
for the wave functions is\footnote{It is important that the integrand in
(\ref{e14}) not contain the complex conjugate of $\psi_m(x)$ because it is not an analytic function of $x$. The integrand of (\ref{e14}) must be analytic because we will need to have the freedom to move the contour of integration in the complex-$x$ plane as the parameter $\delta$ in the Hamiltonian (\ref{e2}) varies. Thus, we do not distinguish between the states $|n\rangle$ and $\langle
n|$.}
\begin{equation}
\int dx\,\psi_m(x)\psi_n(x)=(-1)^n\,\delta_{mn}.
\label{e14}
\end{equation}

In terms of $Z_n$ the expression for the momentum-space two-point Green's function is 
\begin{equation}
G_2(E) = \sum_{n=1}^\infty { Z_n \over E^2+M_n^2},
\label{e15}
\end{equation}
where $M_n=E_n-E_0$. This formula is the quantum mechanical analog of the K\"allen-Lehmann representation \cite{BAR} for the two-point Green's function.

The wave-function renormalization constants $Z_n$ must satisfy the general constraint \begin{equation} \sum_{n=1}^\infty Z_n=1. \label{e16} \end{equation} To derive this result we evaluate the sum by substituting (\ref{e11}): \begin{eqnarray} \sum_{n=1}^\infty Z_n &=& \sum_{n=1}^\infty 2(E_n-E_0)[\langle0|\phi(0)|n\rangle]^2\nonumber\\
&=& \sum_{n=0}^\infty 2(E_n-E_0)[\langle0|\phi(0)|n\rangle]^2\nonumber\\
&=& \sum_{n=0}^\infty\left[\langle0|\phi(0)H|n\rangle\langle n|\phi(0)|0\rangle -\langle0|H\phi(0)|n\rangle \langle n|\phi(0)\right.|0\rangle\nonumber\\
&&\qquad\qquad+\langle0\left.|\phi(0)|n\rangle \langle n|H\phi(0)|0\rangle
- \langle0|\phi(0)|n\rangle \langle n|\phi(0)H|0\rangle\right]\nonumber\\
&=& \langle0|\left[\phi(0)H\phi(0)-H\phi(0)\phi(0)+\phi(0)H\phi(0)-\phi(0)\phi
(0)H\right]|0\rangle,
\label{e17}
\end{eqnarray}
where we have assumed the completeness property $1=\sum_{n=0}^\infty |n\rangle \langle n|$ for the states $|n\rangle$. We then reduce (\ref{e17}) using the equal-time commutation relation of the field $\phi$: \begin{eqnarray} \sum_{n=1}^\infty Z_n=\langle0|[[\phi(0),H],\phi(0)]|0\rangle=\langle0|i[
\dot\phi(0),\phi(0)]|0\rangle=\langle0|i(-i)|0\rangle=\langle0|0\rangle=1.
\label{e18}
\end{eqnarray}

We can calculate the two-point Green's function using several different kinds of perturbative methods. For example, we can sum the one-particle irreducible Feynman graphs having two external legs. Alternatively, we can expand the wave functions and energies in Rayleigh-Schr\"odinger perturbation series and then use (\ref{e13}) and (\ref{e15}). For a one-dimensional conventional $g\phi^4$ theory whose Hamiltonian is \begin{equation} H={1\over2}p^2+{1\over2}m^2x^2+gx^4\qquad(g\geq0),
\label{e19}
\end{equation}
the Euclidean-space Feynman rules are $-24g$ for a vertex and $1\over E^2+m^2$ for a line. The results for the first few wave function renormalization constants are \begin{eqnarray} Z_1 &=& 1-{9\over 8}\epsilon^2+{297\over8}\epsilon^3-{138405\over128}\epsilon^4
+{1015137\over32}\epsilon^5-{991058565\over1024}\epsilon^6+{7942058469\o
+ver256}
\epsilon^7+\cdots,\nonumber\\
Z_3 &=& {9\over8}\epsilon^2-{297\over8}\epsilon^3+{138105\over128}\epsilon^4-
{2015199\over64}\epsilon^5+{974210895\over1024}\epsilon^6-{30738579111\over1024}
\epsilon^7+\cdots,\nonumber\\
Z_5 &=& {75\over32}\epsilon^4-{15075\over64}\epsilon^5+{8419425\over512}
\epsilon^6-{1027873125\over1024}\epsilon^7 + \cdots,\nonumber\\ Z_7 &=& {2205\over256}\epsilon^6-{222705\over128}\epsilon^7+\cdots,
\label{e20}
\end{eqnarray}
where $\epsilon=g/m^3$. For this theory the $Z_{2n}$ vanish by parity. Evidently, the constraint (\ref{e16}) is satisfied perturbatively: \begin{equation} Z_1+Z_3+Z_5+Z_7=1+{\rm O}(\epsilon^8). \label{e21} \end{equation}

What happens for the non-Hermitian $-g\phi^4$ theory that is obtained when we set $\delta=2$ in (\ref{e2})? A perturbative calculation in powers of $g$ gives exactly the same results as above except that $\epsilon$ must be replaced by $-\epsilon$ in (\ref{e20}). Thus, the perturbation expansions now do not alternate in sign. As a result, one might think that each of the $Z_{2n+1}$ would have an imaginary part because the series in (\ref{e20}) are divergent. (That is, one might expect there to be a cut on the positive axis in the complex-$\epsilon$ plane.) However, this is not the case. Indeed, it follows easily from (\ref{e13}) that $Z_n$ is real for all $n$. The key point is that the wave functions are all eigenstates of ${\cal PT}$ \cite{X3}, and thus exhibit the symmetry $[\psi(x)]^* = \psi(-x)$.

Operationally, the reason that $Z_n$ is real is that there is an exponentially small nonperturbative contribution that exactly cancels the discontinuity across the cut. This contribution occurs in all the $Z_n$, so it is no longer true that $Z_{2n}=0$.\footnote{One must be careful to avoid drawing wrong conclusions on the basis of perturbative calculations. For example, a perturbative calculation of the one-point Green's function $G_1$ gives $0$ to all orders in powers of the coupling constant $g$. However, the correct result is nonzero, pure imaginary, and exponentially small \cite{Ya}: $G_1\sim -i K m^{3/2}g^{-2/3}\exp[-m^3/(3g)]$ as $g\to0^+$, where $K$ is a positive constant. This nonperturbative result is due to a soliton solution of the classical field equations.}

Using Runge-Kutta we have found the numerical solution to the Schr\"odinger equation (\ref{e12}) and computed the first eight $Z_n$ for a $-g\phi^4$ quantum mechanical theory using (\ref{e13}). We find that the $Z_n$ alternate in sign with $Z_1$ being positive. When $\epsilon=g/m^3=0.01$, we can see the nonperturbative behavior of $Z_{2n}$; these numbers are nonzero and exponentially small: \begin{equation} \begin{array}{ll}
Z_1 \qquad &  \,\,\,\,\,0.9998757421               \\
Z_2 \qquad &  -0.2341326909\times 10^{-17}\\
Z_3 \qquad &  \,\,\,\,\,0.1649676955 \times 10^{-3} \\
Z_4 \qquad &  -0.6881796661\times 10^{-15}\\
Z_5 \qquad &  \,\,\,\,\,0.8686216150\times 10^{-7} \\
Z_6 \qquad &  -0.8216102337\times 10^{-12}    \\
Z_7 \qquad &  \,\,\,\,\,0.2227856879\times10^{-9}  \\
Z_8 \qquad &  -0.1027951293\times 10^{-10}\\
\end{array}
\label{e24}
\end{equation}
Apparently, the $Z_{2n}$ increase until they reach a maximum and then decay to zero. The sum of the renormalization constants is unity in accordance with (\ref{e16}).

When $\epsilon=0.1$, the partial sums of the $Z_n$ approach $1$ in an oscillatory fashion: \begin{equation} \begin{array}{lrc} n\qquad & Z_n\qquad & \qquad \sum_{k=1}^n Z_k \\
1\qquad &   1.1793780652 & \qquad 1.1793780652 \\
2\qquad &  -0.2117900432 & \qquad 0.9675880220 \\
3\qquad &   0.0362453240 & \qquad 1.0038333461 \\
4\qquad &  -0.0042084418 & \qquad 0.9996249043 \\
5\qquad &   0.0004074167 & \qquad 1.0000323210 \\
6\qquad &  -0.0000348628 & \qquad 0.9999974582 \\
7\qquad &   0.0000027315 & \qquad 1.0000001897 \\
8\qquad &  -0.0000001996 & \qquad 0.9999999901 \\
\end{array}
\label{e23}
\end{equation}
It is difficult to prove completeness when the Hamiltonian is not Hermitian, but the fact that the numerical sum of the renormalization constants is unity is powerful evidence that the states are complete. This is the first time that direct evidence for the completeness of the states of $H$ in (\ref{e2}) has been presented.\footnote{Indirect evidence for completeness in the massless ($m=0$) case was found in Ref.~\cite{X9}.}

In conclusion we point out that many of the results we have found here for one-dimensional Euclidean space immediately extend to the theory characterized by the Hamiltonian density (\ref{e1}) in $D$-dimensional Euclidean space. In particular, we can show that the $2n$-point Green's functions are all real and that the $(2n+1)$-point Green's functions are all pure imaginary. This result follows from the ${\cal PT}$ symmetry of the path-integral representation of the Green's functions. A detailed study of the wave function renormalization constant for $D>1$ will be discussed in a future paper.

\section*{ACKNOWLEDGMENTS}
\label{s1}
This work was supported by the U.S.~Department of Energy.

\end{document}